\documentclass[twocolumn,prl,floatfix,superscriptaddress]{revtex4}

\usepackage{verbatim}
\usepackage{amsmath}
\usepackage{amssymb}
\usepackage{graphicx}
\usepackage{hyperref}
\usepackage{multirow}
\usepackage{array}
\usepackage{float}

\begin{document}

\title{Hard Gap in Epitaxial Semiconductor-Superconductor Nanowires}

\author{W.~Chang}

\affiliation{Center for Quantum Devices, Niels Bohr Institute, University of Copenhagen, Copenhagen, Denmark}

\affiliation{Department of Physics, Harvard University, Cambridge, Massachusetts 02138, USA}

\author{S.~M.~Albrecht}

\affiliation{Center for Quantum Devices, Niels Bohr Institute, University of Copenhagen, Copenhagen, Denmark}

\author{T.~S.~Jespersen}

\affiliation{Center for Quantum Devices, Niels Bohr Institute, University of Copenhagen, Copenhagen, Denmark}

\author{F.~Kuemmeth}

\affiliation{Center for Quantum Devices, Niels Bohr Institute, University of Copenhagen, Copenhagen, Denmark}

\author{P.~Krogstrup}

\affiliation{Center for Quantum Devices, Niels Bohr Institute, University of Copenhagen, Copenhagen, Denmark}

\author{J.~Nyg{\aa}rd}

\affiliation{Center for Quantum Devices, Niels Bohr Institute, University of Copenhagen, Copenhagen, Denmark}

\affiliation{Nano-Science Center, Niels Bohr Institute, University of Copenhagen, Copenhagen, Denmark}

\author{C.~M.~Marcus}

\affiliation{Center for Quantum Devices, Niels Bohr Institute, University of Copenhagen, Copenhagen, Denmark}

\maketitle

{\bf Many present and future applications of superconductivity would benefit from electrostatic control of carrier density and tunneling rates, the hallmark of semiconductor devices. One particularly exciting application is the realization of topological superconductivity \cite{Kitaev} as a basis for quantum information processing \cite{DasSarmaTopoComp, JasonAliceaTopoComp}. Proposals in this direction based on proximity effect in semiconductor nanowires are appealing because the key ingredients are currently in hand \cite{DasSarmaMajorana1D, VonOppenMajorana1D}.  However, previous instances of proximitized semiconductors show significant tunneling conductance below the superconducting gap, suggesting a continuum of subgap states---a situation that nullifies topological protection \cite{DasSarmaTopoProtection, DanielLossPoisoning}. Here, we report a hard superconducting gap induced by proximity effect in a semiconductor, using epitaxial Al-InAs superconductor-semiconductor nanowires. The hard gap, along with favorable material properties and gate-tunability, makes this new hybrid system attractive for a number of applications, as well as fundamental studies of mesoscopic superconductivity. }

Key signatures of topological superconductivity, including a characteristic zero-bias tunneling peak appearing at finite magnetic field, have been reported by several groups over the past few years \cite{LeoMajorana, MotyMajorana, MingtangMajorana, HughMajorana}. In all cases, a soft gap is also seen, indicated by sizable subgap conductance.  The origin of the soft gap is not fully understood, with recent theory attributing it to disorder at the semiconductor/superconductor interface \cite{DasSarmaSoftGap}.  Besides complicating an already complex mesoscopic system by allowing alternative (Kondo) processes that themselves can give rise to zero-bias tunneling peaks, subgap states are fatal to topological protection. This is because quasiparticles occupying subgap states will inadvertently participate in braiding, thus influencing resulting quantum states in an unpredictable and possibly time-dependent way \cite{DasSarmaTopoProtection, DanielLossPoisoning}. 

\begin{figure}[b]
\center \label{figure1}
\includegraphics[width = 8.2cm]{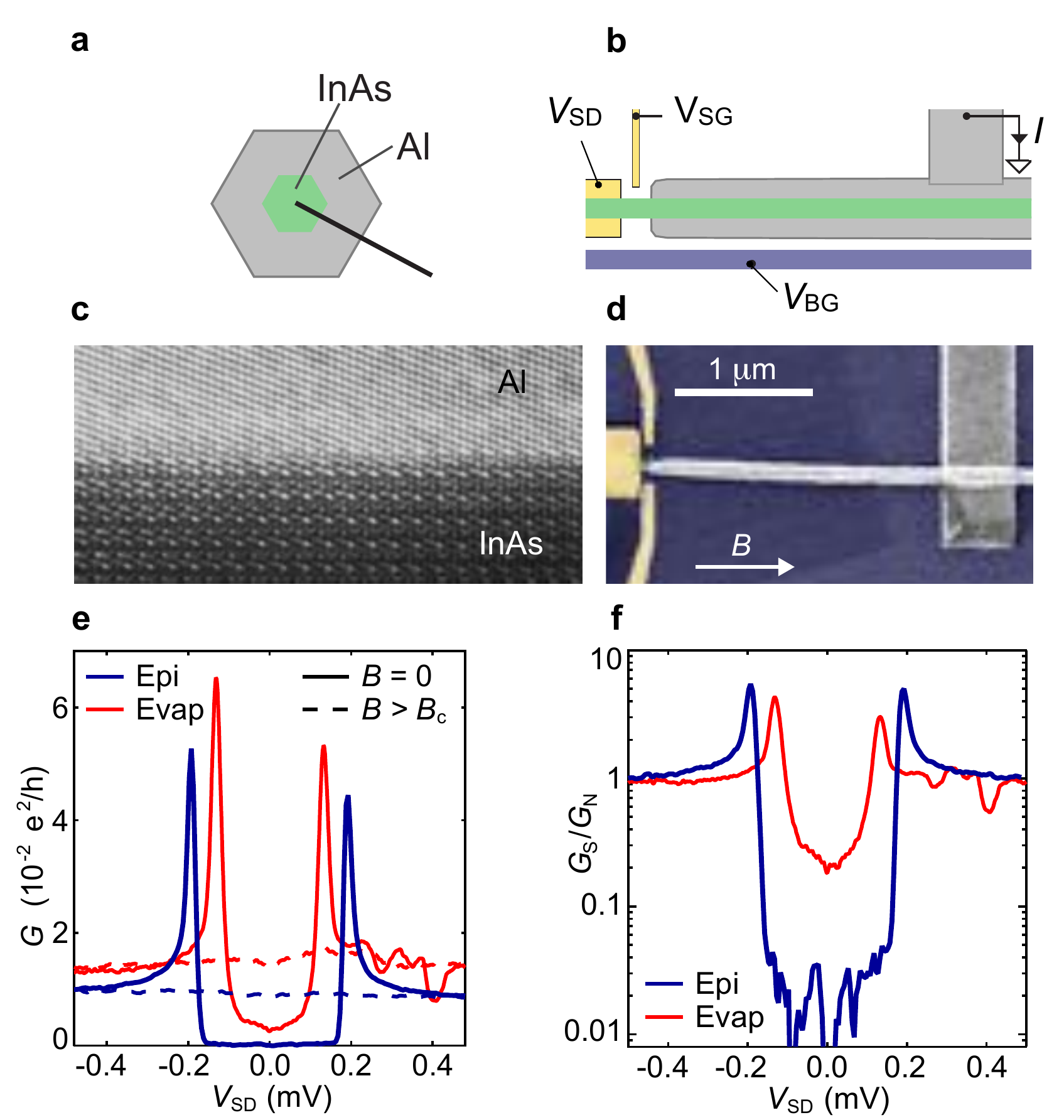}
\caption{\footnotesize{\textbf{Epitaxial full-shell device and hard induced gap.} \textbf{a}, Schematic cross section of epitaxial full-shell nanowire with InAs core (green) and Al shell (gray). \textbf{b}, Measurement setup, showing Ti/Au leads (yellow), InAs nanowire (green), Al shell (gray). \textbf{c}, Transmission electron micrograph of epitaxial N-S interface along the cut in (a). \textbf{d}, Scanning electron micrograph of lithographically similar device (false color). \textbf{e}, Differential conductance as a function of source-drain voltage of an epitaxial full-shell device (blue) and an evaporated control device (red) at $B = 0$ (solid) and above the critical field $B>B_{c}$ (dashed). \textbf{f}, Normalized differential conductance. Epitaxial full-shell nanowires exhibit a subgap conductance suppression by a factor of $\sim$ 100. 
}}
\end{figure}

InAs nanowires were grown in the wurtzite [0001] direction by molecular beam epitaxy (MBE) using gold nanoparticles as catalysts \cite{PeterInAs}. After the nanowires reached a length of 5--10~$\mu$m, Al was grown at low temperature by angled deposition within the MBE chamber.  The resulting semiconductor-superconductor interface, shown in Fig.~1c, appears coherent, domain matched, and impurity free.  Material growth is described in detail in Ref.~\cite{KrogstrupGrowth}. Rotating the substrate during Al growth results in full-shell nanowires with epitaxial interfaces on all facets (Fig.~1a); directional growth without rotating yields half-shell nanowires, with epitaxial Al on two or three facets of the hexagonal InAs core (Fig.~5a). The nanowires were dispersed onto a doped Si substrate with a 100~nm oxide. The Al shell was contacted by superconducting Ti/Al (5/130~nm) and the InAs core (exposed with a selective Al etch) with normal Ti/Au (5/80~nm). Modest \textit{in situ} ion milling was used to improve contact between both the core and shell to leads. A device similar to the one measured is shown in Fig.~1d. Control devices were fabricated by etching away the Al shell and evaporating Ti/Al in select areas (Figs.~1b and 1d). (The 5~nm Ti sticking layer improved gap hardness in all control devices. See supplementary information). 

Measurements were carried out in a dilution refrigerator with a base temperature of 20~mK. Carrier density in the exposed InAs was tuned via the backgate voltage, $V_{\mathrm{BG}}$; the side gate was not used in these measurements. External magnetic field, $B$, was applied along the nanowire axis, unless stated otherwise.  Seven epitaxial devices (along with eight control devices) have been measured to date and show similar behavior.

Tunneling spectra of a full-shell epitaxial device and an evaporated control device in the weak tunneling regime, with conductance of the exposed core tuned to $G \ll G_0 = 2e^2/h$, are shown in Fig.~1e. In the superconducting state ($B = 0$), differential conductance $G_{\mathrm{S}}$ as a function of source-drain voltage, $V_{\mathrm{SD}}$, showed strongly suppressed conductance between symmetric peaks. Above a critical value of field, $B_{\mathrm{c}}$ ($\sim$75~mT for epitaxial, $\sim$250~mT for control), both devices showed featureless normal-state tunneling conductances, $G_{\mathrm{N}}$, of $\sim 0.01\, e^2/h$. Ratios  $G_{\mathrm{S}}/G_{\mathrm{N}}$ for the epitaxial and control devices are shown in Fig.~1f.
The positions of the peaks in $G_{\mathrm{S}}$ indicate an induced gap of $\Delta^*=$~190~$\mu$eV, similar to the gap of bulk Al. Figure 1f shows the subgap conductance suppressed by a factor of  $\sim100$ relative to either normal state ($B > B_{\mathrm{c}}$) or above-gap conductance. The evaporated control device shows a slightly smaller induced gap of 140~$\mu$eV, and a suppression of subgap conduction by a factor up to $\sim 5$, comparable to previous measurements in proximitized InAs and InSb nanowires \cite{LeoMajorana, MotyMajorana, MingtangMajorana, HughMajorana, WillyABS, FinckMajorana}. 

\begin{figure}[t!]
\center \label{figure1}
\includegraphics[width = 8.2cm]{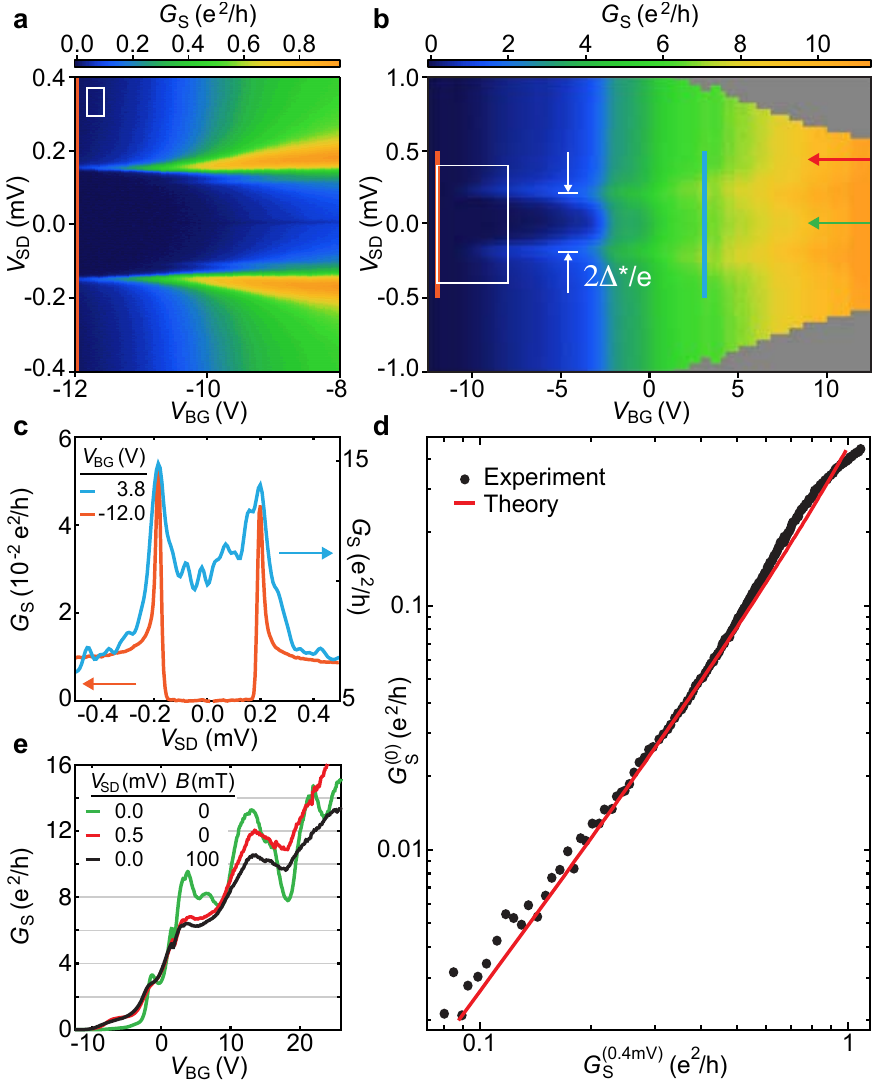}
\caption{\footnotesize{\textbf{Gate dependence of conductance of full-shell device.} \textbf{a}, \textbf{b}, Differential conductance, $G_{\mathrm{S}}$, of a full-shell device as a function of backgate voltage, $V_{\mathrm{BG}}$, and source-drain voltage, $V_{\mathrm{SD}}$. \textbf{c}, Vertical cuts of \textbf{a} and \textbf{b} in the tunneling (orange), and high-conductance (blue) regimes. \textbf{d}, Zero-bias conductance, $G_{\mathrm{S}}$,  as a function of high-bias conductance, measured at $V_{\mathrm{SD}}$ = 0.4 mV (black circles), along with theory of $G_{\mathrm{S}}$ as a function of $G_{\mathrm{N}}$ from Ref.~\cite{BeenakkerNSJunction} with no adjustable parameters (red curve). \textbf{e}, Conductance as a function of $V_{\mathrm{BG}}$ at zero-bias, above-gap bias, and normal state ($B=$ 100 mT $>B_{\mathrm{c}}$) shows plateaus at unexpected values.
}}
\end{figure}

Increasing $V_{\mathrm{BG}}$ in the full-shell device increased both subgap and above-gap conductance (Fig.~2). Conductance peaks at $V_{\mathrm{SD}} =$~190~$\mu$V, indicating the induced gap, did not depend on gate voltage. At positive gate voltages (more open barrier),  subgap conductance exceeds the corresponding normal state conductance, as expected for a moderate-transmission barrier \cite{BTK, BeenakkerNSJunction}. Enhanced subgap conductance is evident in Fig.~2c, which shows two vertical cuts taken at low and high backgate voltages (orange and blue lines in Figs.~2a and 2b). In Fig.~2d, the superconducting zero-bias conductance is plotted as a function of above-gap conductance ($V_{\mathrm{SD}}=$ 0.4~mV) along with theoretical dependence of $G_{\mathrm{S}}(V_{\mathrm{SD}}=0)$ on $G_{\mathrm{N}}(V_{\mathrm{SD}}=0)$ \cite{BeenakkerNSJunction}
\begin{equation}
G_{\mathrm{S}}\big|_{V_{\mathrm{SD}}=0} = 2G_0\frac{G_{\mathrm{N}}^2}{(2G_0-G_{\mathrm{N}})^2},
\end{equation}
with no fit parameters. Using the high-bias conductance ($V_{\mathrm{SD}} =$~0.4~mV~$>\Delta^*/\mathrm{e}$) in place of the normal state conductance is justified by their observed equality in the experiment (see Fig.~1f). Agreement between experiment and the one-channel limit of theory \cite{BeenakkerNSJunction} over a broad range of conductances indicates that transmission in the constriction is single channel.

The device reported in Fig.~2 exhibited conductance steps as a function of $V_{\mathrm{BG}}$ (Fig.~2e), a typical signature of quantum point contacts (QPC). Zero-bias conductance in the normal state (black line) shows plateaus at values close to 1, 3, 6, and 10 $e^2/h$. These unconventional quantization values could be attributed to imperfect transmission of one-dimensional conduction modes \cite{ballisticInAs} or symmetries in the transverse confining potential of the nanowire \cite{unevenQPC}. In addition, while we have subtracted line resistances from our measurement set-up, we cannot independently determine contact resistances within the device, which affect plateau values. In the superconducting state and at source-drain bias above $\Delta^*/e$ (red line), the device conductance shows a similar behavior, but begins to deviate above 6 $e^2/h$. Plateaus are less well defined at zero-bias in the superconducting state (green line). Instead, conductance oscillates around the normal state values and peaks on the lower $V_{\mathrm{BG}}$ edge of the normal state plateaus. 

\begin{figure}[t!]
\center \label{figure3}
\includegraphics[width = 8.2cm]{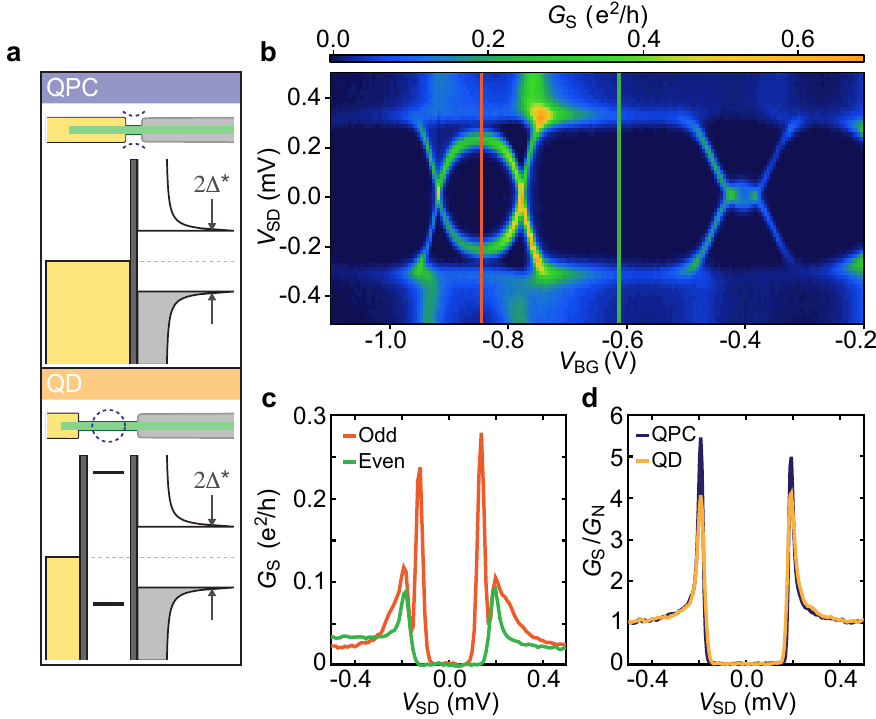}
\caption{\footnotesize{\textbf{Comparing quantum point contact and quantum dot devices.} \textbf{a}, Schematic of tunneling from normal metal lead to proximitized wire via quantum point contact (QPC) barrier (upper panel) and quantum dot (QD) barrier (lower panel). \textbf{b}, Andreev bound states in a quantum dot appear as subgap conductance features. Cuts through a Coulomb valley with odd (orange) and even (green) occupancy. \textbf{c},  Tunneling spectra along cuts in \textbf{b}. \textbf{d}, Comparison of tunneling measurements of QPC and QD devices in an even valley. 
}}
\end{figure}

In some devices, both epitaxial and evaporated, the exposed core region forms a quantum dot (QD) rather than a QPC (Fig.~3a). The formation of a QD versus a QPC barrier depend on the length of the exposed wire, as indicated in Fig.~3(a), but is not yet under full experimental control. In the normal state of an epitaxial full-shell  device with a QD barrier, Coulomb blockade diamonds and Kondo related even-odd structures can be identified (see supplementary information). Since the charging energy of the reported QD device is larger than the induced gap $\Delta^*$, when $V_{\mathrm{BG}}$ is tuned to the middle of an even Coulomb diamond, the discrete QD states are far from the edge of the induced gap. The QD thus acts as a single tunnel barrier between the normal lead and the proximitized InAs core. Accordingly, tunneling spectra for QPC and even-valley QD devices were found to be essentially identical (Fig.~3d).  

In odd-occupied Coulomb valleys, symmetric subgap resonances (SGRs) were observed, forming a characteristic eye shape (Fig.~3b). These SGRs, arising from Andreev bound states or Yu-Shiba-Rusinov states \cite{Yu, Shiba, Rusinov}, crossed due to Coulomb interaction, have been previously investigated experimentally and theoretically \cite{SeigoABS,SilvanoSpinSplit, WillyABS, NadyaMasonABS, PilletABS, TobiasMengABS, SeigoKondoABS, YeyatiABS}. Similar QD structure and their associated SGRs in the superconducting state are also observed in the evaporated-Al control devices. Vertical cuts at the particle-hole symmetry point of an odd (orange) and even (green) Coulomb valley are shown in Fig.~3c.

\begin{figure}[t!]
\center \label{figure4}
\includegraphics[width = 8.2cm]{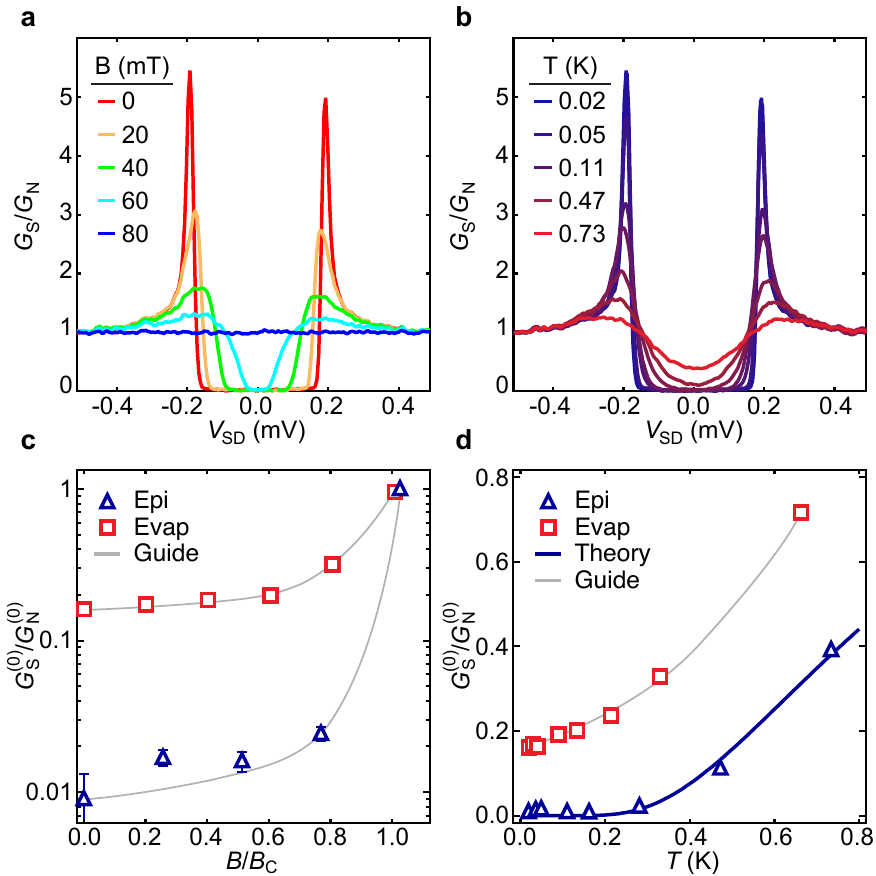}
\caption{\footnotesize{\textbf{Magnetic field and temperature dependence of induced gaps.} \textbf{a}, Magnetic field dependence of full-shell device. \textbf{b}, Temperature dependence of full-shell device. \textbf{c, d}, Normalized zero-bias conductances of epitaxial device (blue triangles) and evaporated (control) device (red squares). Blue line in \textbf{d} is a fit to theory, Eq.~(2), with one fit parameter (see text).}}
\end{figure}

Figure 4 shows the evolution of the proximity-induced gap as a function of magnetic field and temperature. We compare the normalized zero-bias conductance of an epitaxial full-shell device against an evaporated control device in Figs.~4c and 4d. Since the subgap conductance in epitaxial devices is close to our experimental noise-floor, we average over a 40 $\mu$V window centered about zero-bias and define this value as $G_\text{S}^{(0)}/G_\text{N}^{(0)}$. For better comparison, we normalize the applied magnetic field by the critical fields of each device in Fig.~4c. Fig.~4d shows the normalized zero-bias conductance as a function of temperature. Temperature dependence of tunneling conductance of a N-S junction is given by the expression \cite{tinkham}
\begin{equation}
\frac{G_\text{S}}{G_\text{N}}\bigg|_{V_{SD} = 0}=\sqrt{\frac{2\pi\Delta^*}{k_BT}}e^{-\Delta^*/k_BT},
\end{equation}
where $k_B$ is the Boltzmann constant and $T$ is the temperature. From the theoretical fit, we extract an induced gap of 160~$\mu$eV, close to, but not identical to the 190~$\mu$eV measured directly from tunneling spectroscopy.  We note in Fig.~4a and 4c that the floor of the induced gap in the epitaxial devices remains close to zero conductance, rising sharply only when $B$ approaches $B_c$. Retaining a hard gap at finite magnetic fields is important for potential applications in topological quantum computing. We do not know of a theoretical treatment of this dependence to which we can compare the data.

Devices with half-shell nanowires were fabricated by identical methods, though with two superconducting Al leads instead of one, both leads contacting the Al half-shell and the InAs core on the uncovered side of the nanowire (Fig.~5). Tunneling spectroscopy on these devices also shows very low subgap conductance and a gap of 180~$\mu$eV (Fig.~5c), slightly smaller than the induced gap in the full-shell devices. Subgap conductance is a factor of $\sim 50$ below the normal state or high-bias conductance, significantly better than in the evaporated control devices, but not quite as low as the full-shell device for reasons that are not yet understood. 

Using two superconducting leads in the half-shell device allowed us to measure the conductance of the nanowire while the electron density in the half-exposed InAs core is tuned via a side or back gate. As illustrated in Fig.~5b, conductance across the two Al leads was measured in a current biased configuration with the device in the normal state ($B_{\perp} =$ 100~mT $> B_c$). Conductance remained roughly constant at $\sim$10 $e^2/h$ below $V_{\mathrm{BG}} \sim 3\,$V, then rose to $\sim$45 $e^2/h$ at more positive  $V_{\mathrm{BG}}$ (Fig.~5e). We interpret the saturated conductance at negative gate voltages to be the conductance of the Al shell, and the subsequent increase in conductance at positive gate voltages as due to a parallel conduction channel through the InAs core. Using the capacitance model from Ref.~\cite{CapacitanceModel}, we estimate the following transport parameters for the InAs core: carrier density (at high $V_{\mathrm{BG}}$), $n=5\times10^{18}~\mathrm{cm}^{-3}$; mobility, $\mu=3300~\mathrm{cm}^2/\mathrm{Vs}$; elastic scattering length, $l_e=100~\mathrm{nm}$. These are typical values for InAs nanowires reported in Refs.~\cite{LeoSInAsS, ThomasInAs}. However, the expected resistance for our Al shell should be on the order of $10~\Omega$. The higher measured resistance could be attributed to additional contact resistance between the Al leads and the Al shell, or disorder in the Al shell for this particular sample. Regardless of series resistance, the observed saturation of conductance at the negative end of the gate voltage range suggests that the wire is fully depleted at that point. Future experiments with multiple side gates will improve control of density along the wire.

\begin{figure}[t!]
\center \label{figure5}
\includegraphics[width = 8.2cm]{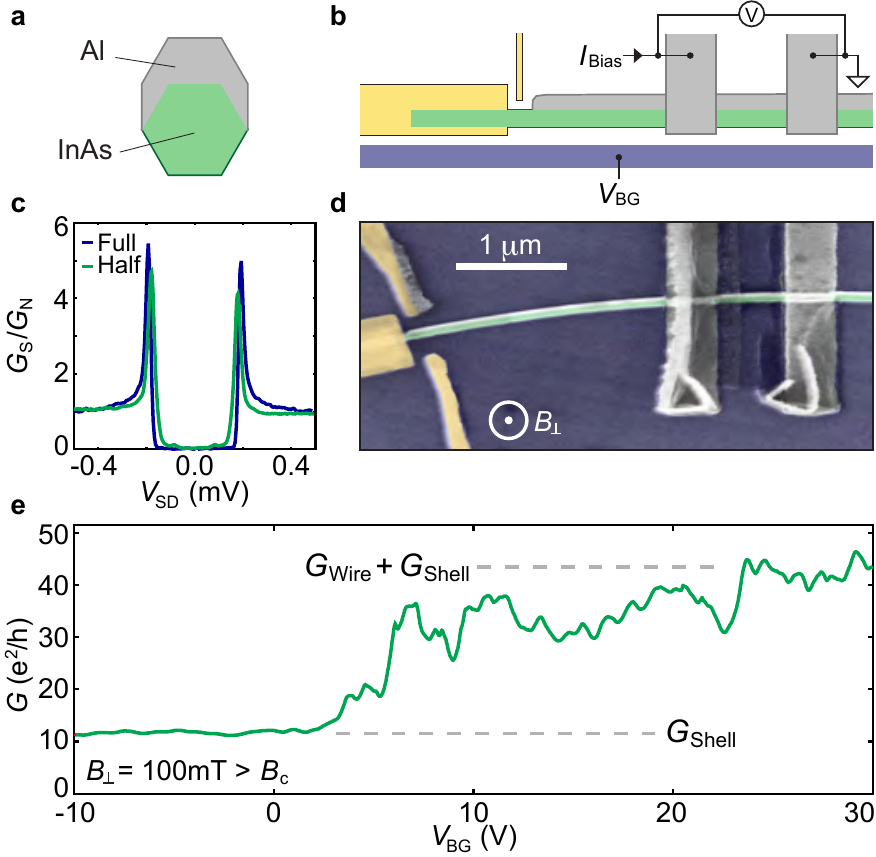}
\caption{\footnotesize{\textbf{Epitaxial half-shell device and gate-tunability of InAs core} \textbf{a}, Cross section illustration of a half-shell nanowire. \textbf{b} Gate-tunability measurement schematic of a half-shell nanowire device. \textbf{c}, Comparison of induced gap quality between an epitaxial full-shell and half-shell device. \textbf{d}, SEM micrograph of a lithographically similar device. \textbf{e}, Conductance of the Al shell and InAs core as a function of $V_{\mathrm{BG}}$. Saturating conductance at negative gate voltage suggests wire is depleted in this regime, with all conductance in this regime due to the shell and contact resistance. 
}}
\end{figure}

While the full-shell nanowires provide fully protective coating as well as an interesting geometry---a cylindrical superconductor---it is presumably the half-shell devices that are of more direct applicability to topological superconductivity and Majorana devices. The possibility of controlling the subband occupation in a large spin-orbit, large g-factor quasi-one-dimensional semiconductor, while maintaining a hard induced superconducting gap, makes the epitaxial half-shell nanowire an ideal platform for nanowire-based Majorana devices and many other applications.

We thank E. Johnson for electron microscopy and K.~Flensberg for valuable discussions. Research supported by Microsoft Project Q, the Danish National Research Foundation, the Carlsberg Foundation, the Villum Foundation, the Lundbeck Foundation, and the European Commission. 

\bibliographystyle{naturemag}

\pagebreak
\onecolumngrid
\setcounter{figure}{0}

\renewcommand{\figurename}{\footnotesize{\textbf{Supplementary Figure}}}
\renewcommand\thefigure{\footnotesize{\textbf{\arabic{figure}}}}

\section{Supplementary Information for ``Hard Gap in Epitaxial Superconductor-Semiconductor Nanowires``''}

\subsection{Additional information on device fabrication}

We sonicated the growth substrate in methanol to liberate the epitaxially grown nanowires. Droplets of the nanowire suspension were then deposited on target substrates and allowed to dry. Nanowires were then optically located relative to pre-fabricated alignment marks. Finally we patterned electrodes onto the substrate with standard electron-beam lithography techniques. 

Aluminum Etchant - Type D, manufactured by Transene Company Inc., was used to remove the Al shell from the InAs core.

Native oxides, both on the Al shell and the exposed InAs core, were removed with Ar ion-milling. This was performed in the same chamber as the electron-beam evaporator used for metals deposition.

\subsection{Control devices}

\subsubsection{Without Ti sticking layer}

\begin{figure}[h!]
\center 
\includegraphics[width=\textwidth]{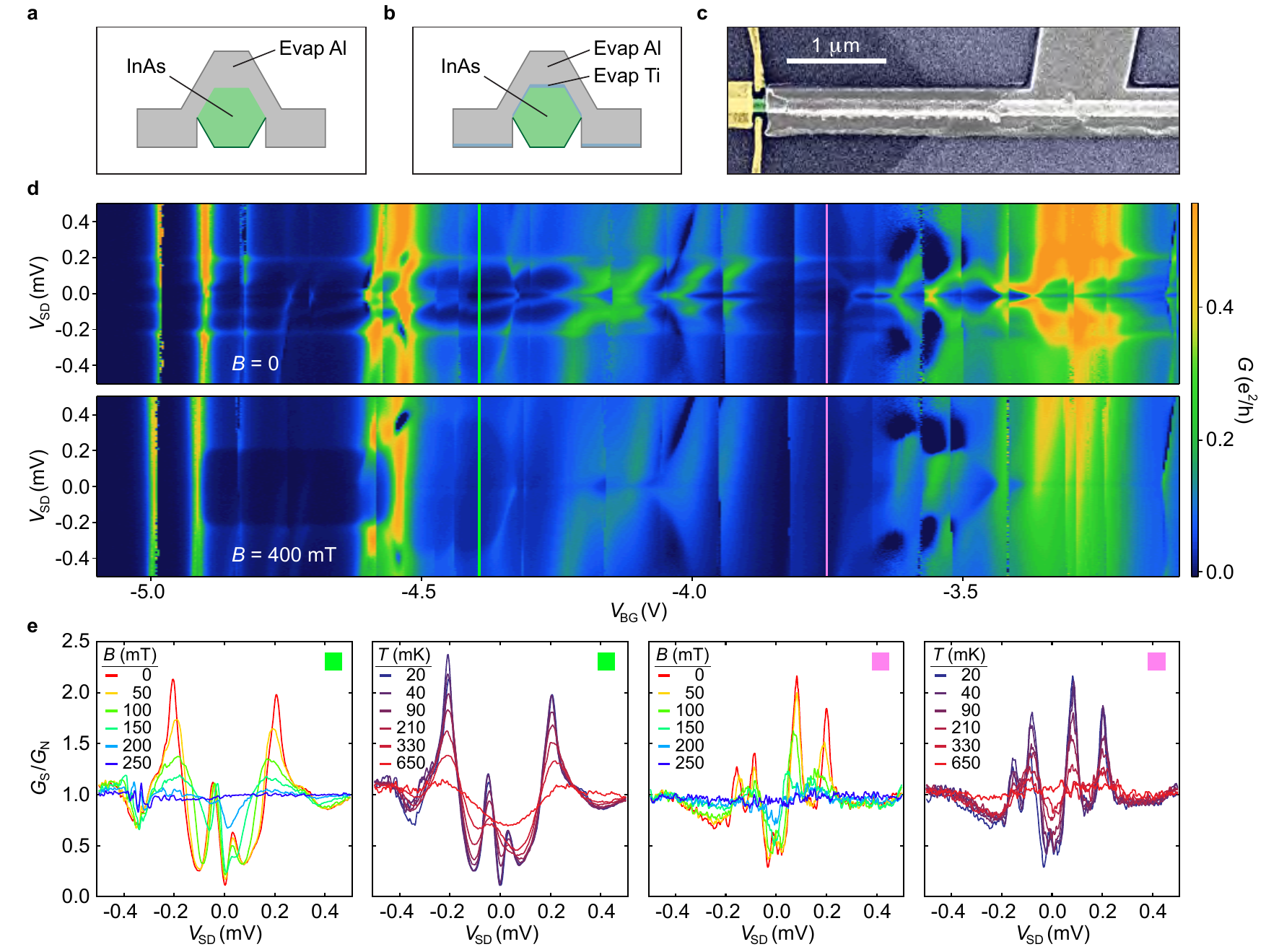}
\caption{\footnotesize{\textbf{Geometry and measurements of evaporate control devices (without Ti sticking layer).} \textbf{a} and \textbf{b}, Cross-section schematic of control devices with evaporated Al film and evaporated Ti/Al film. \textbf{c}, False color SEM micrograph of a lithographically similar control device. \textbf{d}, Differential conductance as a function of $V_{\mathrm{SD}}$ and $V_{\mathrm{BG}}$. \textbf{e}, Vertical cuts of \textbf{d} at various magnetic fields and temperatures. 
}}
\end{figure}

Epitaxial half-shell nanowires from the same growth were used for both the epitaxial half-shell devices and the evaporated control devices. For the control, all of the native Al shell was chemically removed. Al, or Ti/Al (Ti being the sticking layer), was then evaporated onto the remnant InAs core to create a final device similar to the one shown in Supplementary Fig.~1c. Supplementary Figs.~1a and 1b illustrate the cross-sectional profile of these control devices.  

In this section we focus only on evaporated control devices without a Ti sticking layer. Supplementary Fig.~1d shows the differential conductance of a control device as a function of $V_{\mathrm{SD}}$ and $V_{\mathrm{BG}}$. The lower panel shows data from the same region in $V_{\mathrm{BG}}$, but taken at $B=$ 400~mT $>B_{\mathrm{c}}$. The device appears to be highly switchy as the tunneling spectrum is discontinuous in $V_{\mathrm{BG}}$. We can see faint suggestions of Coulomb diamond structures, but the lack of a clear even-odd structure tells us that there are potentially multiple ill-defined QDs in the InAs core. In the superconducting state, there is a backgate-independent induced gap below $|V_{\mathrm{SD}}|\sim$~200~$\mu$V. Populating the device tunneling spectrum are numerous SGRs. The gap and the SGRs originate from the superconducting proximity effect since they disappear at magnetic fields above $B_{\mathrm{c}}$.

At no point in $V_{\mathrm{BG}}$ of this device are we able to avoid the SGRs. This makes extracting the minimum normalized sub-gap conductance difficult. Our best attempts are shown in  Supplementary Fig.~1e, at backgate voltages indicated by the vertical green and pink lines in Supplementary Fig.~1d. We show the evolution of the tunneling spectrum as a function of magnetic field and temperature. In these examples, the normalized sub-gap conductance suppression is at best a factor of 5. Four evaporated control devices without Ti sticking layers were measured, and all of them showed similar behavior. 

\subsubsection{With Ti sticking layer}

\begin{figure}[h!]
\center \label{S2}
\includegraphics[width=\textwidth]{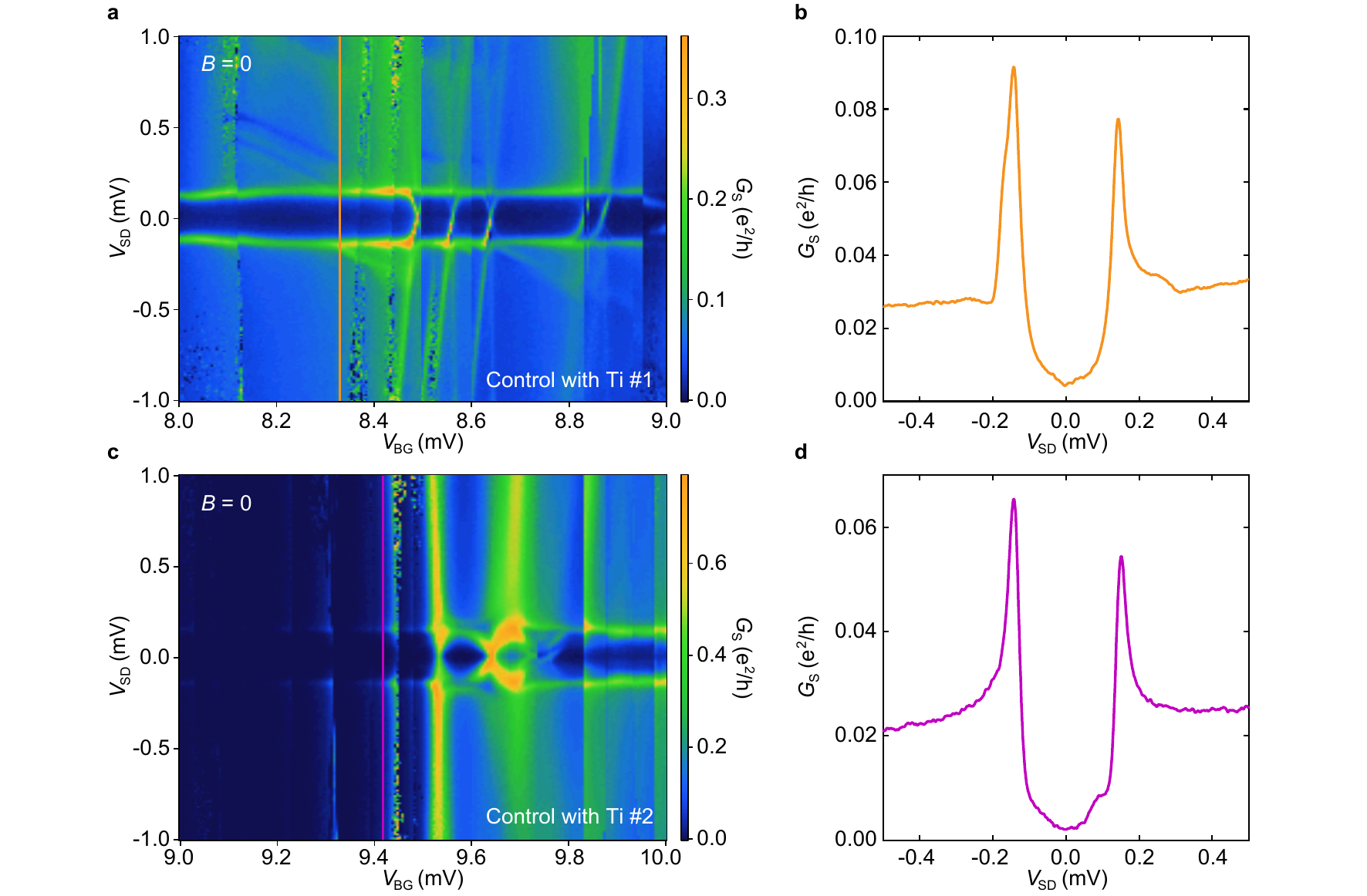}
\caption{\footnotesize{\textbf{Tunneling spectrum of two evaporated control devices with a Ti sticking layer.} \textbf{a} and \textbf{c}, Tunneling spectrums as a function of $V_{\mathrm{SD}}$ and $V_{\mathrm{BG}}$ of control devices \#1 and \#2. \textbf{b} and \textbf{d}, Induced gap measurements of the evaporated control devices taken at $V_{\mathrm{BG}}$ values indicated by the orange and purple lines in \textbf{a} and \textbf{c}.
}}
\end{figure}

Supplementary Fig.~2 shows the tunneling spectrum of two evaporated control devices with Ti sticking layers. Both devices are switchy (discontinuities in $V_{\mathrm{BG}}$), but compared to devices without a Ti sticking layer, it is now possible to move away from the SGRs and extract a minimum normalized sub-gap conductance. It is also possible to identify odd and even Coulomb valleys by the SGRs and the Kondo resonances. Supplementary Figs.~2b and 2d show conductance traces at $V_{\mathrm{BG}}$ values indicated by the orange and purple lines in Supplementary Figs.~2a and 2c.

\subsection{Epitaxial full-shell QPC device}

\begin{figure}[h!]
\center \label{S3}
\includegraphics[width=\textwidth]{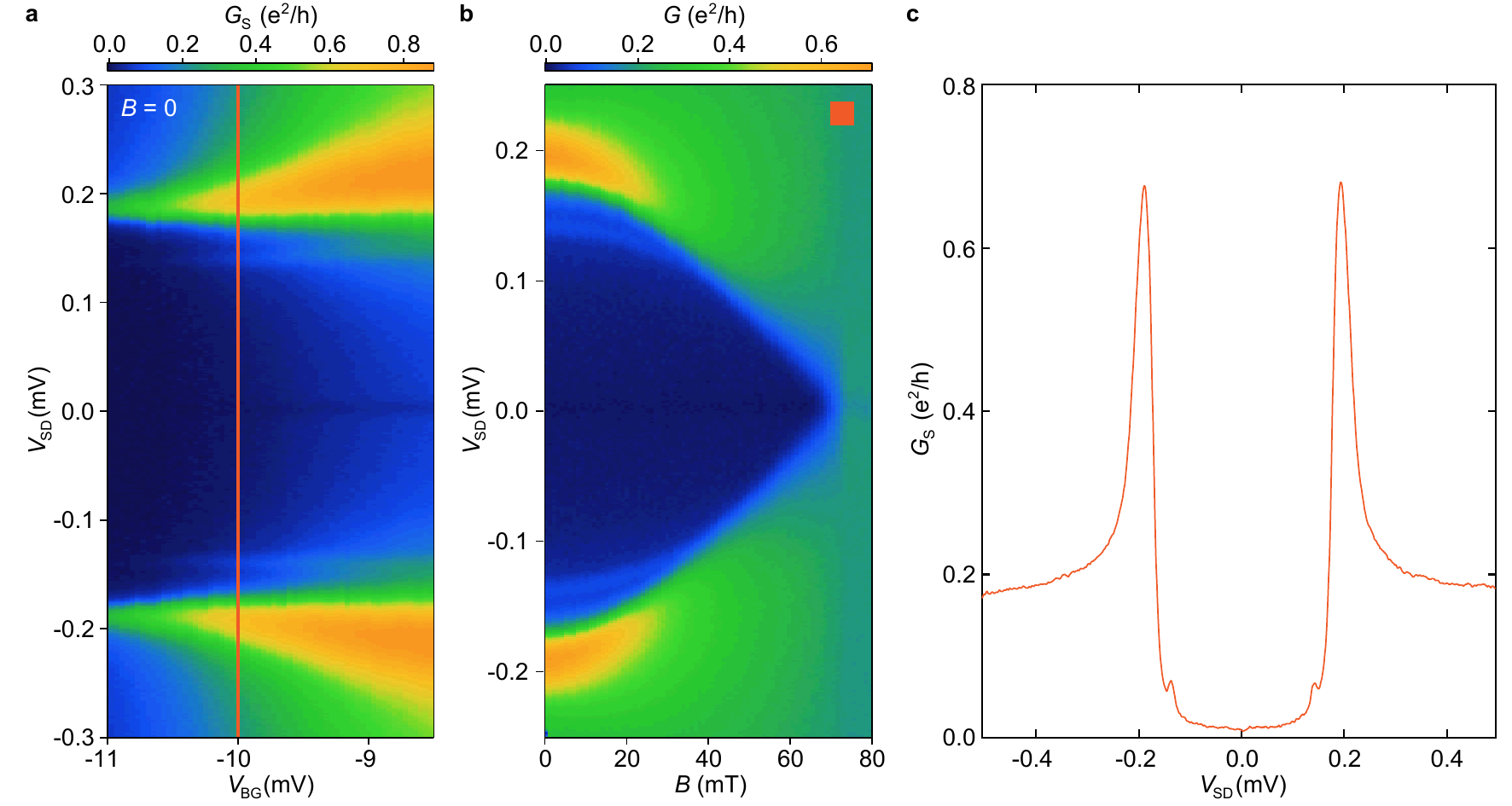}
\caption{\footnotesize{\textbf{ABS of the first sub-band in the epitaxial full-shell QPC device.} \textbf{a}, Tunneling spectrum below the first conductance plateau. A pair of horizontal SGRs can be seen at the edge of the induced gap. \textbf{b}, Magnetic field dependence of the SGRs. \textbf{c}, Vertical cut of \textbf{a} (orange line), showing two small conductance peaks at the edge of the induced gap. 
}}
\end{figure}

Additional data on the single ABS in the epitaxial full-shell QPC device is shown in Supplementary Fig.~3. Data shown in Supplementary Fig.~3a is identical to data shown in Fig.~2a in the main text. Here, we change the aspect ratio to place emphasis on the ABS. Supplementary Fig.~3b shows the evolution of the induced gap and the ABS as a function of magnetic field. The ABS remains close to the gap edge, then merges into the continuum above $B\sim$~40~mT. It shows up most prominently as two peaks in an one-dimensional trace taken at $V_{\mathrm{BG}}=$~-10~V (Supplementary Fig.~3c).

\subsection{Epitaxial full-shell QD device}

\begin{figure}[h!]
\center \label{S4}
\includegraphics[width=\textwidth]{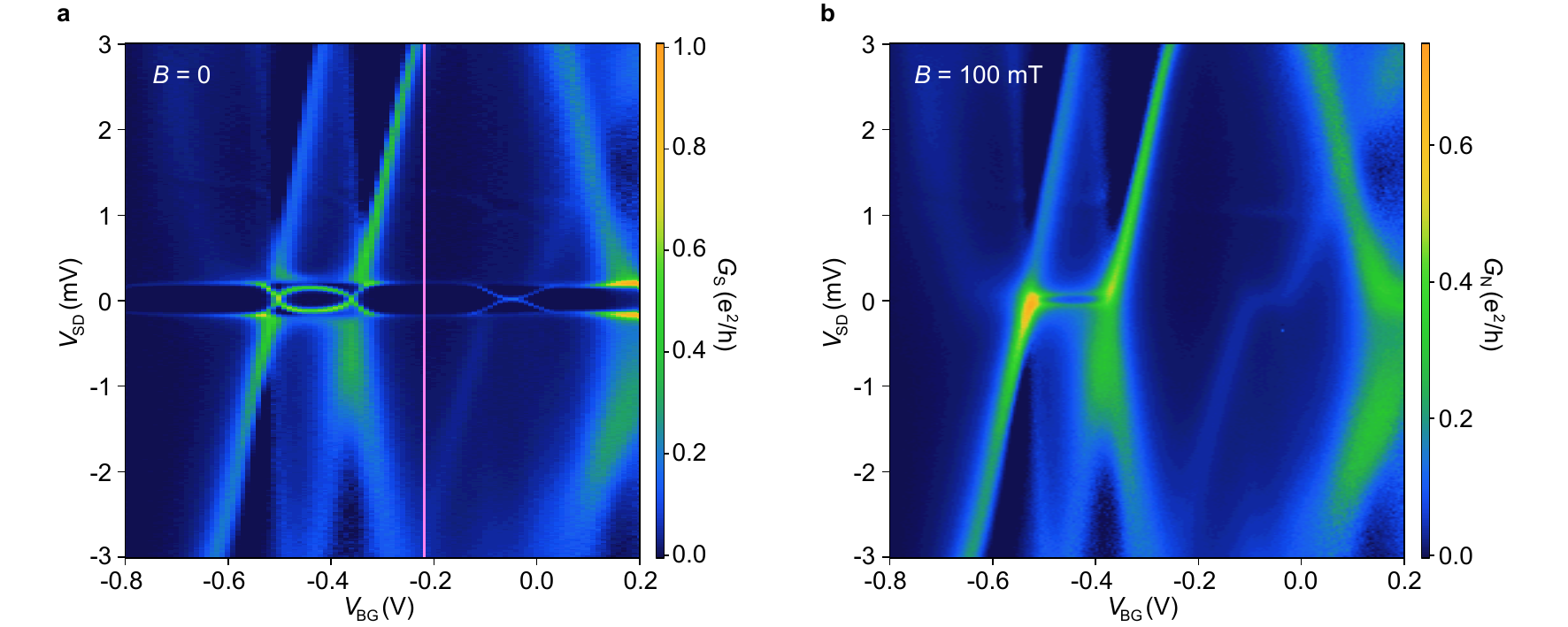}
\caption{\footnotesize{\textbf{Charge stability diagram of an epitaxial full-shell QD device.} \textbf{a} and \textbf{b}, Differential conductance as a function of $V_{\mathrm{SD}}$ and $V_{\mathrm{BG}}$ at $B=0$ and $B=$~100~mT respectively. Pink line in \textbf{a} indicates the backgate voltage at which the induced gap measurements shown in Figs.~3c and 3d (main text) is taken. 
}}
\end{figure}

We turn our attention to measurements from epitaxial full-shell QD devices. Distinct Coulomb diamond resonances with charging energy on the order of 1~meV can be seen in Supplementary Fig.~4. Furthermore, even-odd structure can be seen from the presence of both Kondo resonances (normal state) and SGRs (superconducting state) in alternating Coulomb diamonds. QD data shown in Fig.~3d (main text) is taken at the particle-hole symmetry point of an even Coulomb diamond (pink line in Supplementary Fig.~4a).

\subsection{Epitaxial half-shell QD device}

\begin{figure}[h!]
\center \label{S5}
\includegraphics[width=\textwidth]{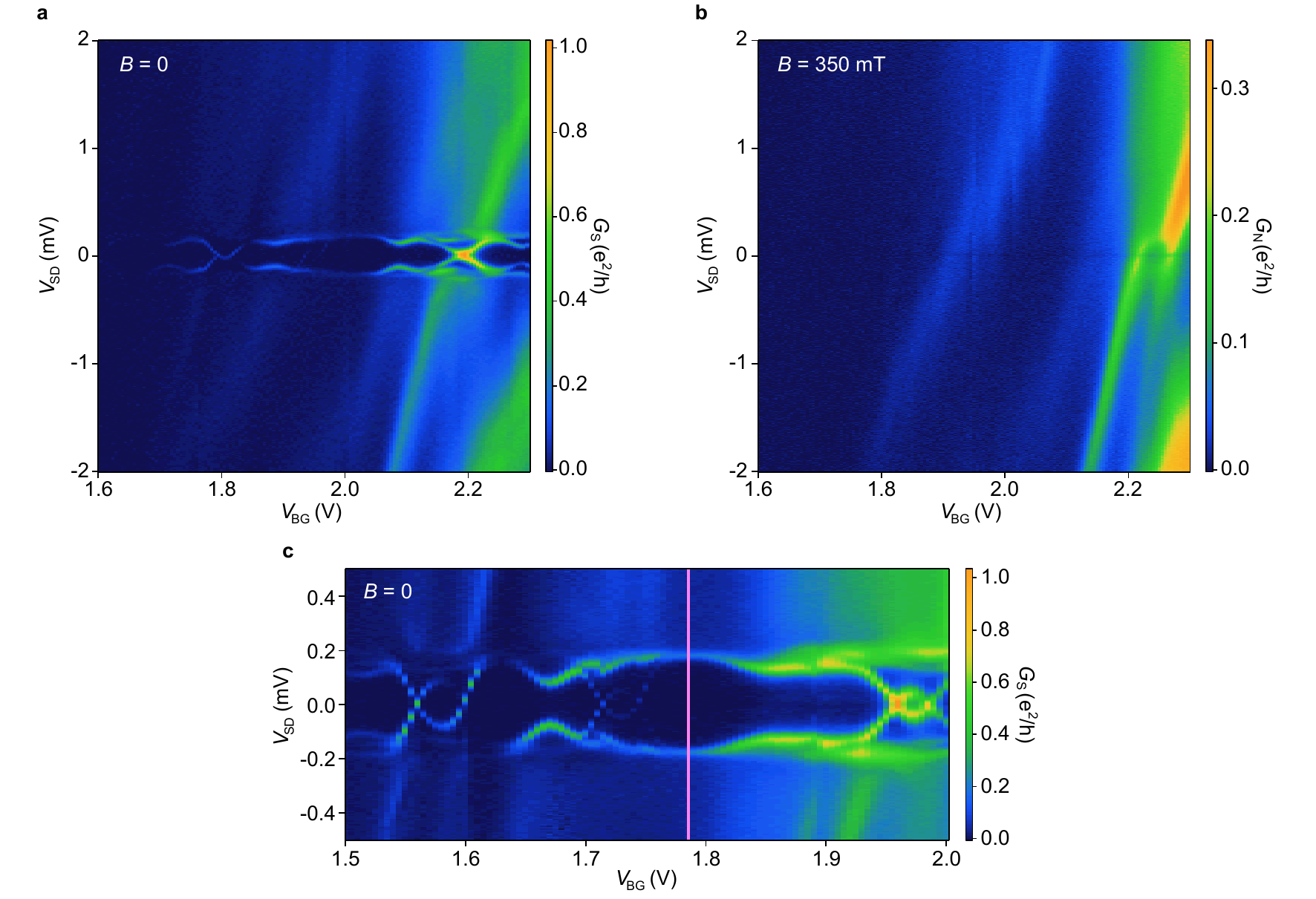}
\caption{\footnotesize{\textbf{Charge stability diagram of an epitaxial half-shell QD device.} \textbf{a} and \textbf{b}, Charge stability diagram of the device at $B=0$ and $B>B_{\mathrm{c}}$. Coulomb diamonds with charging energy on the order of 1~meV are evident. In the normal state, a spin-split Kondo resonance occurs near $V_{\mathrm{BG}}=$~2.2~V. \textbf{c}, Close-up of the sub-gap spectrum in the backgate voltage range of \textbf{a} and \textbf{b}; numerous SGRs are visible. The pink line indicates the backgate voltage value at which the induced gap measurement shown in Fig.~5c (main text) is taken. 
}}
\end{figure}
 
 Three epitaxial half-shell QD devices were measured. In Supplementary Figs.~5a and 5b we show the charge stability diagrams of one of the devices at zero and finite magnetic fields. The QD nature of this device is evident from Coulomb diamonds, and its charging energy is on the order of 1~meV. In the normal state (Supplementary Fig.~5b), we see a pair of spin-split Kondo resonances near $V_{\mathrm{BG}}=$~2.2~V. In the superconducting state, the Kondo resonance turns into a pair of SGRs (Supplementary Fig.~5a). To extract a measurement of the induced gap, we move away from the SGRs in $V_{\mathrm{BG}}$. Data shown in Fig.~5c (main text) is taken at a backgate voltage value indicated by the pink line in Supplementary Fig.~5c. 
 
\end{document}